\documentclass{article}
\usepackage{amssymb}
\usepackage{amsmath}
\usepackage{boxedminipage}

\begin{document}

\bigskip

\begin{center}
\large La stabilit\'e du Kink interne dans les configurations h\'elico{\"\i}dales de grand rapport \\ d'aspect $hx \ll 1$
\end{center}

\begin{center}
{\sc Jean-Louis Soul\'e} \qquad Mars 1980 \footnote {
J'ai trouv\'e les notes ci-dessous dans les papiers de mon
p\`ere apr\`es son d\'ec\`es. Elles ont \'et\'e tap\'ees
\`a l'Institut des Hautes Etudes Scientifiques, Bures-sur-Yvette.
Je remercie G.Laval et M.Vittot pour leurs conseils
et la lecture de ce texte.  

\hfill Christophe Soul\'e, soule@ihes.fr}\\
Association Euratom-CEA, CEN-FAR
\end{center}

\bigskip
\begin{center}
\bf Abstract
\end{center}

We get the general expression for the stability criterion 
with respect to the internal kink under helicoidal
discharges, if one assumes that the helicity is finite and
the aspect ratio is large ($hx \ll 1$).

We show that the helicity relative to the cylindric geometry
has a stabilizing effect. This effect can be greater than
the toric effect, when the helicoidal pitch is smaller
that the toric radius.

\bigskip

\begin{center}
\bf R\'esum\'e
\end{center}

\medskip

On \'etablit l'expression g\'en\'erale du crit\`ere de stabilit\'e relatif au kink interne pour les d\'echarges h\'elico{\"\i}dales, dans l'hypoth\`ese d'une h\'elicit\'e finie et d'un grand rapport d'aspect ($hx \ll 1$). 

\smallskip

On montre l'effet stabilisant de l'h\'elicit\'e par r\'ef\'erence \`a la g\'eom\'etrie  cylindrique. Cet effet peut \^etre plus important que l'effet torique, lorsque le pas h\'elico{\"\i}dal est plus petit que le rayon torique.

\bigskip

\begin{center}
\bf Introduction
\end{center}

\medskip

La stabilit\'e du kink interne a d\'ej\`a \'et\'e \'etudi\'ee (dans le cadre de la MHD id\'eale) pour diff\'erentes configurations, en consid\'erant comme un petit param\`etre l'\'ecart \`a la g\'eom\'etrie droite infinie. On a ainsi trait\'e la g\'eom\'etrie droite finie et circulaire [], la g\'eom\'etrie droite infinie non circulaire [], la g\'eom\'etrie torique finie []. Dans toutes ces situations, il s'agit d'\'equilibres poss\'edant une invariance par d\'eplacement. Dans ce cas il restait \`a traiter la g\'eom\'etrie h\'elico{\"\i}dale de pas fini (et grand), ce qui est l'objet de ce travail. Pour l'essentiel, la m\'ethode utilis\'ee reste la m\^eme que pr\'ec\'edemment []; elle sera rappel\'ee au fur et \`a mesure de son emploi. Pr\'ecisons que nous nous contenterons d'\'etablir l'expression de l'\'energie propre du mode, le traitement de la couche et le calcul du taux de croissance restant identique \`a celui de la configuration torique (v\'erifier $\xi_{\varphi}$).

\bigskip

\begin{center}
\bf Les coordonn\'ees h\'elico{\"\i}dales
\end{center}

\medskip

(Indiquer les notations de Mercier.)
 A partir du syst\`eme direct de coordonn\'ees polaires $r,\varphi ,z$ on forme le syst\`eme :
$$
r=r \qquad u = h \varphi + \ell z \qquad v = -\ell \varphi + hz
$$
o\`u les constantes $h$ et $\ell$ sont normalis\'ees par $h^2 + \ell^2 = 1$ (avec $h > 0$; $\ell > 0$ ou $\ell < 0$). Chaque h\'elice $(H)$ d\'efinie par $r = {\rm cte}$, $v = {\rm cte}$ sera un lieu d'invariance pour l'\'equilibre ($z$ est normalis\'e par $r_0$). Sa transform\'ee rotationnelle est $\frac{h}{\ell}$. Elle a pour p\'eriode en $z$ $2\pi \frac{\ell}{h}$, en $u$ $\frac{2\pi}{h}$. $(H)$ est droite pour $\ell$ positif, gauche pour $\ell$ n\'egatif (axe magn\'etique). La n\'ecessit\'e d'introduire des pr\'ecisions ({\it cf.} Annexe A) provient essentiellement du fait que le syst\`eme $r,u,v$ n'est pas orthogonal. On utilisera une base de vecteurs orthogonaux, obtenue en rempla\c cant $\vec{\nabla} u$ par un vecteur $\vec U = \vec{\nabla} u - K \vec{\nabla} u$ tangent \`a $H$. La difficult\'e est alors en partie report\'ee et se retrouve dans le fait que $\vec U$ n'est pas un gradient : on a ${\rm rot} \, \vec U = \frac{2h\ell \, \vec U}{h^2 + \ell^2 r^2}$. Pour simplifier au maximum les calculs, il y a int\'er\^et \`a effectuer deux autres modifications :

\medskip

\noindent 1) renormaliser la variable $r$ pour simplifier la m\'etrique. On posera $x = \int_1^r \vert \nabla v \vert \, dr$, ce qui donne :
$$
d\ell^2 = \frac{1}{\vert \nabla v \vert^2} \, (dx^2 + dv^2) + \left( \frac{\vec U \cdot \vec{d\ell}}{U} \right)^2
$$

\noindent 2) introduire des coordonn\'ees et des vecteurs complexes :
\begin{eqnarray}
&\displaystyle S = \frac{x-iv}{2} \quad t = \frac{x+iv}{2} &\vec S = \vec{\nabla} x + i \, \vec{\nabla} v \nonumber \\
&&\vec T = \vec{\nabla} x - i \, \vec{\nabla} v \nonumber
\end{eqnarray}

\bigskip

\begin{center}
\bf Equilibres h\'elico{\"\i}daux
\end{center}

\medskip

Dans le syst\`eme de coordonn\'ees $x,u,v$, l'\'equation du flux polo{\"\i}dal s'\'ecrit :
$$
V^2 LF + 2h\ell \, U^2 f (F) + f \, \frac{df}{dF} + \frac{1}{U^2} \, \frac{dp}{dF} = 0 \leqno (1)
$$
o\`u
$$
LF = \frac{\partial^2 F}{\partial x^2} + \frac{\partial^2 F}{\partial v^2} - h^2 \, \frac{U}{v^2} \, \frac{\partial F}{\partial x} \quad V^2 LF = \frac{1}{U^2} \, {\rm div} (U^2) \sim \Delta U \, . \leqno (2)
$$

\medskip

\noindent {\bf Solution G\'en\'erale.} Le champ\footnote{Le champ de r\'ef\'erence $B=U$ comporte un courant (sans force). On pourrait prendre un champ de r\'ef\'erence sans courant (par exemple $B = \nabla u$) mais la pr\'esence d'une composante polo{\"\i}dale impliquerait les choses.} se mettant sous la forme : 
$$
\vec B = f(F) \, \vec U + \vec U \times \vec{\nabla} F \leqno (3)
$$
on choisit l'unit\'e de longueur pour que l'axe magn\'etique corresponde \`a $r=1$ ($x=0$) et $v=0$. 

\smallskip

On supposera dor\'enavant que $\vert x \vert$ et $\vert v \vert$ restent petits dans la zone concern\'ee $\vert x \vert \sim \vert v \vert = 0 (\varepsilon) \ll 1$. Ainsi $U$, $V$ et $r$ resteront voisins de l'unit\'e. L'ordering est standard. On a : $x$ et $v$ d'ordre $\varepsilon$, $f$ d'ordre 1, $F$ d'ordre $\varepsilon^2$ (donc $\nabla F$ et $B_p$ d'ordre $\varepsilon$) $\frac{dp}{dF}$ d'ordre 1, $\frac{df}{dF}$ d'ordre 1. Les d\'erivations en $x$ ou $v$ abaissent les quantit\'es d'un ordre. Au point de vue signe, $f$ est positif, $F$ est d\'ecroissant \`a partir de l'axe. Les lignes magn\'etiques sont des h\'elices gauches pour $q>0$ ({\it cf.} Annexe D). Une perturbation \`a $k > 0$ est une h\'elice gauche. Finalement $\frac{h\ell}{k}$ est n\'egatif lorsque les lignes de champ tournent dans le m\^eme sens que l'axe magn\'etique.

\smallskip

On peut d\'eterminer aux deux premiers ordres (Annexe B) des familles d'\'equilibres \`a ``cercles'' d\'ecentr\'es (ce ne sont pas r\'eellement des cercles mais des ellipses parce que $x$ et $v$ ne sont pas exactement des longueurs). [(Autre {\it ordering}, {\it cf.} Annexe B.) En fait $\left.\begin{matrix} \vert x \vert \, h < 1 \\ \vert v \vert \, h < 1 \end{matrix} \right\vert h\ell x$ suffit ({\it cf.} les termes correctifs de l'\'equilibre) d'o\`u ordering des termes non circulaire de l'\'equilibre.]

\smallskip

A partir de l'expression du d\'ecentrement, on peut d\'eduire celle de la modulation du champ polo{\"\i}dal le long d'une surface magn\'etique. Aux deux premiers ordres on a :
$$
\vert B_p \vert \sim \vert B_p \vert_0 \cdot [ 1 + \lambda \, h^2 \rho \cos \theta ]
$$
o\`u $p$ et $\theta$ sont les coordonn\'ees polaires de la section et
$$
\lambda + 1 = \int_0^1 z^3 dz \, \{ Q^2 - (4 \delta \, Q \} + \int_0^1 2z \, dz \, P \, .
$$
On a $q = \frac{\rho f}{B_p}$, $z = \frac{\rho}{\rho_1}$, $\rho_1 , q_1$ et $p_1$ \'etant les valeurs de $\rho$, $q$ et $p$ sur la surface consid\'er\'ee $\frac{p-p_1}{B_p^2 / 2} = P$, $Q = \frac{q_1}{q}$, $h\ell q_1 = \delta$.

\smallskip

Pour la suite, on introduira le coefficient de modulation ``r\'eduit'' d\'efini par :
$$
\tilde\lambda = \lambda + \frac{3}{4} + \delta \ \bigl \Vert \ \tilde\lambda = \int_0^1 z^3 dz \{ Q^2 - 1 - 4\delta (Q=1)\} + \int_0^1 2z \, dz \, P
$$
$\tilde\lambda$ est nul si $p$ et $q$ sont constants.

\bigskip

\begin{center}
\bf Coordonn\'ees intrins\`eques
\end{center}

\medskip

On peut mettre le champ d'un \'equilibre donn\'e h\'elico{\"\i}dal sous la forme
$$
\vec B = \nabla \, [ \varphi - q(F) \, \chi] \times \nabla F
$$
o\`u $\chi$ est ind\'ependant de $\varphi$ et multivoque $(0,2\pi)$ [Hamada]. De $\vec U = \vec{\nabla} u - K \vec V$ on tire $B = \nabla u \times \nabla F + f \vec U - K \vec V \times \vec{\nabla} F$. Donc on doit avoir [si $u=\varphi$] $-\nabla (q\chi) \times \nabla F = f\vec U - K \vec V \times \vec{\nabla} F$. En multipliant par $\vec U$ :
$$
B_p \nabla \vert q \chi \vert = f \, U^2 + K \, V \cdot B_p = {\mathcal J} = B \, \nabla \varphi \, .
$$
Dans le Jacobian ${\mathcal J}$, $KV \cdot B_p$ est du 2$^{\rm e}$ ordre devant $fU^2$. [Le chouia $fU^2$ correspond au fait que $u$ n'est pas au d\'epart exactement \'egal \`a la coordonn\'ee d'Hamada $\varphi$ telle que $\chi = \varphi - q\theta$.] Dans le syst\`eme de coordonn\'ees intrins\`eques $F,\varphi , \chi$ on a :
$$
B \nabla \psi = {\mathcal J} \left( \frac{\partial \psi}{\partial \varphi} + \frac{1}{q} \, \frac{\partial \psi}{\partial \chi} \right)
$$
on a $d\tau = \frac{dF \, d\varphi \, d\chi}{\nabla F \cdot \nabla \varphi \times \nabla \chi}$ avec $\nabla F \cdot \nabla \varphi \times \nabla \chi = \nabla F \cdot U \times \nabla \chi$. D'o\`u $d\tau = - \frac{q}{{\mathcal J}} \, dF \cdot d\mu \, d\chi$. Si on remplace $F$ par $\mu$ d\'efini par $-\frac{q}{f} \, dF = \mu \, d\mu$, on obtient : $d\tau = \frac{f}{{\mathcal J}} \, \mu \, d\mu \, du \, d$. L'\'el\'ement de longueur de l'h\'elice est $\frac{du}{U}$, donc l'\'el\'ement d'aire orthogonale \`a $\vec U$ est $\frac{Uf}{{\mathcal J}} \, \mu \, d\mu \, d\chi$.

\smallskip

Dans le cas des \'equilibres \`a cercles d\'ecentr\'es, on peut obtenir aux deux premiers ordres la formule de changement de coordonn\'ees $x,u,v \to \mu , u , \chi$. Il est de la forme : $x - iv = \Delta (\mu) + \mu \, e^{-i\omega}$ [$\omega = \omega (\mu , \chi)$] on en d\'eduit $dx \, dv = (1 + \Delta' \cos \omega) \, \mu \, \frac{d\omega}{d\chi} \, d\mu \, d\chi$. [Il faut passer de $s,t$ \`a $z , \bar z$, $z = \mu \, e^i \, \chi$.] Comme on a $d\sigma = \frac{dx \, dv}{V^2} = \frac{Uf}{{\mathcal J}} \, \mu \, d\mu \, d\chi$, on a :
$$
(1+\Delta' \cos \omega) \, \frac{d\omega}{d\chi} = \frac{Uf}{{\mathcal J}} \, V^2 \to \omega = \chi + (h^2 \lambda + 2h^2 - \ell^2) \, \mu \sin \chi
$$
$$
2S = x-iv = \mu \, e^{-i\chi} + \frac{\mu^2}{2} \, (h^2 \lambda + 2h^2 - \ell^2) (e^{-2i\chi} - 1) + \Delta (\mu) \, .
$$

\bigskip

\begin{center}
\bf Rappel du probl\`eme
\end{center}

\medskip

Comme pour les configurations pr\'ec\'edemment trait\'ees, un petit param\`etre $\varepsilon$ (le rapport d'aspect inverse) caract\'erise l'\'ecart de la configuration consid\'er\'ee au cylindre circulaire infini. Pour $\varepsilon$ nul, l'instabilit\'e est marginale, et il s'agit de d\'eterminer, \`a l'ordre significatif en $\varepsilon$, l'expression du r\'eservoir d'\'energie $\delta w$.

\smallskip

Comme dans les autres cas, cet ordre est en $\varepsilon^4$ par rapport \`a l'\'energie -- positive -- associ\'ee \`a une perturbation arbitraire. Puisque nous ne cherchons pas \`a obtenir la perturbation elle-m\^eme, nous pouvons profiter des propri\'et\'es variationnelles de $\delta w$ pour limiter la recherche de la fonction d'essai \`a un d\'eveloppement en $\varepsilon$ moins pouss\'e. Si la perturbation minimisante exacte est $\tilde\xi$, si la fonction d'essai est de la forme $\tilde\xi + \delta \xi$, l'erreur sur $\delta w$ sera $\delta w (\delta \xi)$. Cette erreur doit \^etre d'ordre sup\'erieur \`a $\varepsilon^4$. Cette condition sera remplie en imposant \`a l'erreur $\xi$ deux contraintes :

\smallskip

1) d'une part les composantes non contraintes de l'erreur $\delta\xi$ seront d'ordre au moins \'egal \`a $\varepsilon^3 \xi_0$ (soit $\varepsilon^6$ au carr\'e)

\smallskip

2) d'autre part les composantes de l'erreur qui entrent dans $\delta w$ par une contribution en $\varepsilon^2$ seront d'ordre au moins \'egal \`a $\varepsilon^2$.

\smallskip

En d'autres termes, si $\tilde\xi$ satisfait l'\'equation d'Euler $F(\tilde\xi) = 0$, on imposera \`a la fonction d'essai $\tilde\xi + \xi$ de satisfaire  $F (\tilde\xi + \xi) = 0 (\varepsilon^3)$.

\smallskip

Une m\'ethode de r\'esolution dite ``altern\'ee'' est expos\'ee dans l'Annexe C. Une m\'ethode plus sp\'ecifique dite ``directe'' est expos\'ee ci-apr\`es.

\newpage

\begin{center}
\bf La m\'ethode directe
\end{center}

\medskip

Son application est limit\'ee au kink interne. Elle consiste \`a construire une fonction d'essai pour le principe d'\'energie qui soit d'embl\'ee exacte jusqu'\`a l'ordre requis. Ceci est possible parce que jusqu'\`a cet ordre elle est essentiellement ind\'ependante de l'\'equilibre et qu'on peut donc s'appuyer sur le cas simple o\`u le champ se r\'eduit \`a $\vec B = \vec U$.

Cette fonction $\vec{\xi}$ sera obtenue comme la somme de trois parties : $\vec{\xi} = \vec{\xi}_0 + \vec{\xi}_1 + \vec{\xi}_2$ (l'indice rappelant le mode dominant dans chacune de ces parties). 

\smallskip

\noindent 1) La partie principale $\vec{\xi}_1$ (qui sera d\'etermin\'ee la premi\`ere) s'annule au-del\`a de la surface r\'esonante ; en de\c c\`a, elle est strictement ind\'ependante de l'\'equilibre (ainsi d'ailleurs que $\vec{\xi}_0$).

\smallskip

\noindent 2) $\vec{\xi}_1$ co{\"\i}ncide \`a l'ordre principal avec le kink interne du cylindre infini circulaire.

\smallskip

\noindent 3) $\vec{\xi}_0$ est une contribution simple de l'harmonique $m=0$ (une composante purement azimutale et uniforme). [$\vec{\xi}_0$ est inutile puisque ${\xi}$ est d\'efini \`a un multiple de $B$ pr\`es. Ah, mais si, $B$ va changer. Mais alors, il faut l'introduire quand on change $\vec B$.]

\smallskip

\noindent 4) $\vec{\xi}_2$ ne contient que l'harmonique $m=2$ (mais $\xi_1$ en compte aussi). Il est plus petit d'un ordre que l'harmonique $m=1$ de $\vec{\xi}_1$. $\vec{\xi}_2$ est adapt\'e pour que $\vec{\xi}$ satisfasse les conditions de continuit\'e sur la r\'esonance et les conditions aux limites. Il d\'epend enti\`erement de l'\'equilibre particulier.

\smallskip

\noindent 5) Chaque partie satisfait approximativement l'\'equation d'Euler, avec un r\'esidu qui est en valeur absolue d'ordre $\varepsilon^2$ (par rapport \`a $B_m \, \frac{\vert \xi \vert}{\varepsilon}$), aussi bien dans la direction azimutale que m\'eridienne.

\smallskip

\noindent 6) $\vec{\xi}_1$ et $\vec{\xi}_0$ seront exprim\'es en coordonn\'ees h\'elico{\"\i}dales, $\vec{\xi}_2$ en coordonn\'ees intrins\`eques.

\bigskip

\begin{center}
\bf Le tenseur de d\'eformation
\end{center}

\medskip

On utilise l'identit\'e alg\'ebrique g\'en\'erale suivante :
$$
Q + J \times \xi = B \cdot D - \nabla (\xi \cdot B)
$$
o\`u le tenseur $D$ est d\'efini comme suit :
$$
D = \nabla \xi + \nabla^I \xi - I \, {\rm div} \, \xi
$$
(en coordonn\'ees cart\'esiennes : $D_{ij} = \frac{d \xi_j}{dx_i} + \frac{d\xi_i}{dx_j} - \delta_{ij} \sum_i \frac{d\xi_i}{dx_i}$). La construction de $\xi$ consistera essentiellement \`a r\'eduire l'ordre de $D$ (le terme de pression ne pr\'esentant pas de difficult\'e). Ainsi $\phi$ sera essentiellement \'egal \`a $-\xi \cdot B$.

\smallskip

On appellera tenseur m\'eridient $D_m$ l'ensemble des composantes purement m\'eridiennes de $D$ et tenseur azimutal $D_u$ l'ensemble des autres composantes ($D = D_m + D_u$). Il sera n\'ecessaire de choisir $\xi$ pour abaisser $D_u$ \`a l'ordre $\varepsilon^3$ et $D_m$ seulement \`a l'ordre $\varepsilon^2$ (il sera plus simple de dire qu'on annule $D_u$ exactement). Pour cela nous calculerons d'abord les composantes de $D$ sur la base $S,U,T$ \`a l'aide de la formule g\'en\'erale ($A$ et $B$ quelconques) :
$$
A \cdot D \cdot B = A \nabla (\xi \cdot B) + B \nabla (\xi \cdot A) - {\rm div} \, [\vec\xi \, A \cdot B] + \xi \cdot A \times {\rm rot} \, B + \xi \cdot B \times {\rm rot} \, A \, .
$$
Nous poserons : $\xi_S = \xi \cdot S$, $\xi_t = \xi \cdot T$, $\xi_u = \frac{i}{k} \, \xi \cdot U$
$$
A \nabla \xi \cdot B + B \nabla \xi \cdot A - A \cdot B \, {\rm div} \, \xi \, .
$$
Appliquons la formule de la trace (valable pour une base orthonorm\'ee) :
$$
\sum D_{ii} + {\rm div} \, \xi = 0 \, .
$$
[A r\'e\'ecrire en coordonn\'ees {\it norm\'ees} et en base {\it norm\'ee}.] Elle donne ici
$$
\frac{U \cdot D \cdot U}{U^2} + \frac{S \cdot D \cdot T}{V^2} + {\rm div} \, \xi = 0
$$
(car $XDX + VDV = SDT$). On obtient pour $D_u$ :
\begin{eqnarray}
UDU &= &2k^2 \, U^2 \, \xi_u - {\rm div} \, (\vec\xi \, U^2) \nonumber \\
\frac{i}{k} \, SDU &= &V^2 \left( \frac{d\xi_u}{dS} - \Delta \, \xi_u \right) - U^2 \left( 1 - \frac{2h\ell}{k} \, U \right) \xi_S \nonumber \\
\frac{i}{k} \, TDU &= &V^2 \left( \frac{d\xi_u}{dt} + L \, \xi_u \right) - U^2 \left( 1 + \frac{2h\ell}{k} \, U \right) \xi_t \, . \nonumber
\end{eqnarray}
Pour $D_m$ on obtient [${\rm rot} \, S = 0$, ${\rm rot} \, T = 0$, ${\rm rot} \, U \ne 0$] :
\begin{eqnarray}
SDS &= &2V^2 \left( \frac{d\xi_S}{dS} - L \, \xi_S \right) \qquad 2V^2 = ST \nonumber \\
TDT &= &2V^2 \left( \frac{d\xi_t}{dt} + L \, \xi_t \right) \nonumber \\
-SDT &= &+ \, h^2 \, U (\xi_S + \xi_t) + 2k^2 \, V^2 \, \xi_u \, . \nonumber
\end{eqnarray}
En \'ecrivant $\vec{\xi} = \vec S \, \frac{\xi_t}{2V^2} + \vec T \, \frac{\xi_S}{2V^2} - ik \, \vec U \, \frac{\xi_u}{U^2}$ on explicite $U \cdot D \cdot U$ sous la forme :
$$
U \cdot D \cdot U = k^2 \, U^2 \, \xi_u - U \, \frac{V^2}{2} \left[ \frac{d}{dS} \left( \frac{U \xi_t}{V^2} \right) + \frac{d}{dt} \left( \frac{U \xi_S}{V^2} \right)\right] + L \, \frac{U^2}{2} \, [\xi_t - \xi_S]
$$
$$
{\rm div} \, \xi = \vec S \, \nabla \left( \frac{\xi_t}{2V^2} \right) + \vec T \, \nabla \left( \frac{\xi_S}{2V^2} \right) + k^2 \, \xi_u \, .
$$
Faux :
$$
{\rm div} \, \xi = V^2 \left[ \frac{d}{dS} + iK \frac{d}{du} \right] \frac{\xi_t}{2V^2} + V^2 \left[ \frac{d}{dt} - iK \frac{d}{du} \right] \frac{\xi_B}{2V^2} + k^2 \xi_u \, .
$$

\bigskip

\begin{center}
\bf La fonction d'essai ($\xi_1$ et $\xi_0$)
\end{center}

\medskip

$\xi_1$ sera choisi parmi les solutions exactes (\`a trois ordres) de l'\'equation d'Euler relative au champ $\vec B_0 = \vec U$. Dans ce cas (qui est sans force mais pas sans courant) la composante $\xi \cdot U$ ne joue aucun r\^ole dans l'\'equation d'Euler. Nous pouvons donc la fixer par la condition :
$$
\xi \cdot U + \phi = 0
$$
ce qui permet d'\'ecire l'\'equation d'Euler :
$$
U \cdot D = 0
$$
(ce n'est pas une restriction r\'eelle : la solution g\'en\'erale s'obtient \`a partir de l\`a en prenant $\phi = - \xi \cdot U$, {\it puis} en redonnant \`a $\xi \cdot U$ une valeur arbitraire). A l'ordre le plus bas (et sans \'ecrire le facteur $e^{iku}$) nous devons imposer $\xi \cdot X = 1$, $\xi \cdot V = -i$, c'est-\`a-dire $\xi \cdot S = 2$, $\xi \cdot T = 0$. D'autre part \`a cet ordre $\phi$ (et par suite $\xi_i \, U = - \phi$) est n\'ecessairement nul ; sinon $Q_u^* \cdot G_u$ contiendrait une constante positive et $\delta w$ ne serait pas minimum ; cela correspondrait \`a la pr\'esence d'un mode $m=0$ de compression qui se propagerait n\'ecessairement jusqu'\`a la fronti\`ere. Par contre, un mode $m=0$ de rotation (sans composante normale) est acceptable jusqu'\`a la r\'esonance : il sera d\'ecrit par $\vec\xi_0$, avec une valeur de $\xi_u$ appropri\'ee est une valeur nulle de $\phi$ (cette d\'ecomposition en $\vec\xi_1$ et $\vec\xi_0$ est indispensable pour clarifier les calculs).

\smallskip

Revenant \`a $\xi_1$, il doit d'abord satisfaire l'\'equation d'Euler pour $\vec B_0$, c'est-\`a-dire :
$$
U \cdot D = 0 \, ,
$$
ce qui donne [$0(\varepsilon^3 \xi)$] :
$$
U \cdot D \cdot U = 0 \qquad S \cdot D \cdot U = 0 \qquad T \cdot D \cdot U = 0 \, .
$$
Pour les \'el\'ements m\'eridiens de $D$, compte tenu du comportement de $\xi_1$ \`a l'ordre le plus bas, ils sont automatiquement d'ordre un. Mais il est n\'ecessaire qu'ils soient d'ordre deux pour que $\xi_1$ soit la solution g\'en\'erale \`a cet ordre. Puisque $K$ est d'ordre un et nul \`a l'origine, on obtient pour $SDS$ et $TDT$ les conditions~:
$$
\frac{d \xi_S}{dS} \, (0) = 0 \qquad \frac{d \xi_t}{dt} \, (0) = 0 \, .
$$
On verra que cet ensemble de conditions fixe $\xi_1$ suffisamment. Il reste \`a consid\'erer $SDT$. D'apr\`es son expression, il sera d'ordre deux si $h^2 + k^2 \, \xi_u = 0$. Cela s'obtiendra en adjoignant \`a $\xi_1$ le terme $\xi_0$ form\'e purement d'une composante azimutale (autrement dit $\xi_0$ contient la modification de $\xi \cdot U$ qui \'etait a priori arbitraire avec $\vec B_0$, et qui est appropri\'ee pour un champ quelconque). [Soit ${\rm div} \, \xi = 0$ sur l'axe.] Donc pour $\vec\xi_0$ :
$$
\xi_{u0} = - \frac{h^2}{k^2} \, e^{iku} \qquad \vec\xi_0 = \frac{ih^2}{k} \, \frac{\vec U}{U^2} \, e^{iku} \, .
$$

\bigskip

\begin{center}
\bf Le d\'eveloppement limit\'e de la solution
\end{center}

\medskip

La solution $\vec\xi_1$ doit satisfaire les trois \'equations :
$$
UDU = 0 \, , \quad SDU = 0 \, , \quad TDU = 0
$$
et les conditions ``initiales'' sur l'axe magn\'etique :
$$
\xi_S = 2 \, , \quad \xi_t = 0 \, , \quad \xi_u = 0 \, , \quad \frac{d \xi_S}{dS} = 0 \, , \quad \frac{d \xi_t}{dt} = 0 \, .
$$
Bien qu'elles ne d\'eterminent pas compl\`etement $\xi_1$, nous verrons que ces conditions suffisent \`a fixer les quantit\'es qui jouent ult\'erieurement.

Une fa\c con simple de proc\'eder pour \'etudier la solution consiste \`a \'eliminer $\xi_S$ et $\xi_t$ et \`a \'ecrire une \'equation pour $\xi_u$ de la forme :
$$
x = S + t \quad \frac{d^2 \xi_u}{dS \, dt} + A(x) \, \frac{d \xi_u}{dS} + B(x) \, \frac{d \xi_u}{dt} + C(x) \, \xi_u = 0 \, .
$$
Les conditions initiales report\'ees sur $\xi_u$ donnent :
$$
\xi_u = 0 \quad \frac{d \xi_u}{dt} = 0 \quad  \frac{d^2 \xi_u}{dt^2} = 0 \quad  \frac{d \xi_u}{dS} = 2 \left( 1 - \frac{2h\ell}{k} \right) \, , \quad  \frac{d^2 \xi_u}{dS^2} = X \, .
$$
Il est \'evident qu'il existe pour $\xi_u$ un d\'eveloppement convenable (en commen\c cant par $ \frac{d^2 \xi_u}{dS \, dt}$), les termes en $S^p$ et $t^p$ (pour $p \geq 3$) pouvant \^etre choisis arbitrairement.

\smallskip

Cependant, pour obtenir les coefficients qui nous int\'eressent, il est un peu plus court de calculer directement $\xi_S$. On verra qu'il est suffisant de calculer sur l'axe :
$$
\frac{1}{2} \, \frac{d}{dt} \, SDS = \frac{d^2 \xi_S}{dt \, dS} - 2 \frac{dL}{dx}
$$
(compte tenu des propri\'et\'es de $\xi_S$, $L = kK$)
$$
-\frac{d}{dS} \, SDT = + \, 2h^2 \, \frac{dU}{dx} + h^2 U \, \frac{d \xi_t}{dS} + 2 \, k^2 \, V^2 \, \frac{d \xi_u}{dS} \, .
$$
L'inconnue fondamentale est donc $\frac{d^2 \xi_S}{dt \, dS}$.

\smallskip

Posons
$$
X = \frac{U}{V^2} \, \xi_S \quad Y = \frac{U}{V^2} \, \xi_t \quad f = U - \frac{2h\ell}{k} \, U^2 \quad g = U + \frac{2h\ell}{k} \, U \quad L = kK \, .
$$
Les trois \'equations liant $\xi_u$, $X$, $Y$ s'\'ecrivent :

\medskip

(A) \quad $\displaystyle \frac{i}{k} \, \frac{SDU}{V^2} = \frac{d\xi_u}{dS} - L \, \xi_u - fX = 0$

\medskip

(B) \quad $\displaystyle \frac{i}{k} \, \frac{SDV}{V^2} = \frac{d\xi_u}{dt} + L \, \xi_u - g \, Y = 0$

\medskip

(C) \quad $\displaystyle \frac{2UDU}{UV^2} = 2k^2 \,  \frac{U}{V^2} \, \xi_u - \frac{dY}{dS} - \frac{dX}{dt} + L (Y-X) = 0$

\medskip

\noindent ($\xi_u (0) = 0$, $Y(0) = 0$, $X(0) = 2$, $(L(0)=0)$).

\smallskip

Une \'elimination presque compl\`ete de $\xi_u$ entre (A) et (B) r\'esulte de la quasi-permutabilit\'e des op\'erateurs $D_S$ et $D_t$ :
$$
\frac{1}{V^2} \, S \, \nabla = \frac{d}{dS} - L \qquad \hbox{et} \qquad \frac{1}{V^2} \, T \, \nabla = \frac{d}{dt} + L \, .
$$
On multiplie (A) par $\frac{d}{dt} + L$, (B) par $\frac{d}{dS} - L$ et on soustrait :

\medskip

(D) \quad $\displaystyle -2 \, \frac{dL}{dx} \, \xi_u - \frac{d}{dt} \, (fX) + \frac{d}{dS} \, (g \, Y) - L (fX + g \, Y) = 0 \, .$

\medskip

On constate maintenant que (C) et (D) contiennent les m\^emes d\'eriv\'ees : $\frac{dX}{dt}$ et $\frac{dY}{dS}$, qu'on peut s\'eparer par combinaison et \'ecrire facilement en $\xi_S$ et $\xi_t$ :

\medskip

(E) \quad $\displaystyle 0 = \frac{d\xi_S}{dt} + \xi_S \left( \frac{V^2}{U} \, \frac{d}{dx} \, \frac{U}{V^2} + L + \frac{1}{2U} \, \frac{df}{dx} \right) - \xi_t \, \frac{1}{2U} \, \frac{dg}{dx} + \xi_u \left( \frac{V^2}{U^2} \, \frac{dL}{dx} - g \, \frac{k^2}{U} \right)$

\medskip

(F) \quad $\displaystyle 0 = \frac{d\xi_t}{dS} + \xi_t \left( \frac{V^2}{U} \, \frac{d}{dx} \, \frac{U}{V^2} - L + \frac{1}{2U} \, \frac{dg}{dx} \right) - \xi_S \, \frac{1}{2U} \, \frac{df}{dx} + \xi_u \left( - \frac{V^2}{U^2} \, \frac{dL}{dx} - f \, \frac{k^2}{U} \right)$

\medskip

\noindent [$\frac{d\xi_S}{dt}$ et $\frac{d\xi_t}{dS}$ sont les bonnes inconnues puisqu'on fixe $\xi_S$ pour $t=0$ et $\xi_t$ pour $S=0$]. Nous avons besoin de d\'eriver (E) par rapport \`a $S$. En fait, il suffit de le faire sur l'axe en substituant \`a $\xi_S$, $\frac{d\xi_S}{dS}$, $\xi_t$, $\frac{d\xi_t}{dS}$, $\xi_u$, $\frac{d\xi_u}{dS}$ les valeurs d\'ej\`a connues. Pour $\frac{d\xi_u}{dS}$ et $\frac{d\xi_t}{dS}$, on les tire de (A) et (F) :
$$
\frac{d\xi_u}{dS} = 2f \qquad \frac{d\xi_t}{dS} = \frac{df}{dx} \, .
$$
On obtient (en faisant $U=V=1$ quand ils ne sont pas d\'eriv\'es) :
$$
\left. 0 = \frac{1}{2} \, \frac{d^2 \xi_S}{dS \, dt} \right]_0 + \frac{d}{dx} \left( \frac{V^2}{U} \, \frac{d}{dx} \, \frac{U}{V^2} + L + \frac{1}{2U} \, \frac{df}{dx} \right) - \frac{1}{4} \, \frac{df}{dx} \, \frac{dg}{dx} + f \left( \frac{dL}{dx} - g \, k^2 \right)
$$
[c'est bien \c ca la m\'ethode la plus rapide (pour abr\'eger, poser $D_S = \frac{d}{dS} - L$, $D_t = \frac{d}{dt} + L$)].
\begin{eqnarray}
0 &= &\frac{1}{4} \, \frac{d}{dt} \, SDS + \frac{d}{dx} \left( \frac{V^2}{U} \, \frac{dU}{dx} + \frac{V^2}{2U} \, \frac{df}{dx} \right) - \frac{1}{4} \, \frac{df}{dx} \, \frac{dg}{dx} \nonumber \\ 
&+ &\frac{d}{dx} \left( \frac{1-V^2}{U} \, \frac{dU}{dx} + V^2 \, \frac{d}{dx} \, \frac{1}{V^2} + \frac{1-V^2}{2U} \, \frac{df}{dx} \right) - fg \, k^2 + \frac{dL}{dx} \, (2 + f) \nonumber
\end{eqnarray}

\begin{eqnarray}
0 &= &\frac{1}{4} \, \frac{d}{dt} \, SDS - h^2 \, \frac{d}{dx} \left[ U \left( \frac{3}{2} - \delta \, U \right) \right] - \left( \frac{dU}{dx} \right)^2 \left( \frac{1}{4} - \delta^2 \, U^2 \right) \nonumber \\ 
&+ &\ell^2 \, \frac{d}{dx} \left[ (U^2 - 1) \left( \frac{3}{2} \, U - \delta \, U^2 \right) + 2 \, U^3 \right] - k^2 (1-\delta^2) + k^2 (3-\delta) \nonumber
\end{eqnarray}

$$
1 - \frac{1}{V^2} = 1 - r^2 \, U^2 = \frac{\ell^2}{h^2} \, (U^2 - 1)
$$

$$
0 = \frac{1}{4} \, \frac{d}{dt} \, SDS + h^4 \left( \frac{3}{2} - 2 \delta + \delta^2 - \frac{1}{4} \right) + k^2 \left( -2 \, \delta^2 - \frac{\delta^2}{4} + \frac{\delta^3}{2} + 3 \, \delta - 1 \right) \, .
$$

On a $\xi_S = \frac{V^2}{Uf} \left( \frac{d}{dS} - L \right) \xi_u$ avec
$$
\left[ 2k^2 \, \frac{U}{V^2} - \left( \frac{d}{dS} - L \right) \frac{V^2}{Ug} \left( \frac{d}{dt} + L \right) - \left( \frac{d}{dt} + L \right) \frac{V^2}{Ug} \left( \frac{d}{dS} - L \right) \right] \xi_u = 0 \, .
$$
Il suffit de remplacer $\xi_u$ par $\left( \frac{d}{dS} - L \right)^{-1} \frac{Uf}{V^2} \, \xi_S$
$$
\frac{1}{2} \, \frac{d^2 \xi_S}{dS \, dt} - \left( \frac{d}{dx} \, \frac{U}{V^2} \right)^2 + \frac{d^2}{dx^2} \, \frac{U}{V^2} + \frac{d}{dx} \left( \frac{1}{2U} \, \frac{df}{dx} \right) - \frac{1}{4} \, \frac{df}{dx} \, \frac{dg}{dx} + (1+f) \, \frac{dL}{dx} - fg \, k^2 = 0 \, .
$$
Par substitution des expressions $U$, $V$, $f$, $g$, $L$ on obtient :
$$
\frac{1}{2} \, \frac{d^2 \xi_S}{dS \, dt} = k^2 \left( 1 - 2 \delta + \frac{9}{4} \, \delta^2 - \frac{\delta^3}{2} \right) - h^4 \left( \frac{5}{4} - 2 \delta + \delta^2 \right)
$$

\medskip

\begin{center}
\begin{boxedminipage}{11cm} $\displaystyle \frac{1}{4} \, \frac{d}{dt} \, SDS = \frac{1}{2} \, \frac{d^2 \xi_S}{dt \, dS} - \frac{dL}{dx} = \left( 1 - \frac{\delta}{2} \right)^2 (1-2\delta) \, k^2 - h^4 \left( \frac{5}{4} - 2\delta + \delta^2 \right)$ \end{boxedminipage}
\end{center}

\medskip

$$
\delta = \frac{2h\ell}{k}
$$
$$
\frac{d}{dS} \, SDT = - \, 4 \, k^2 (1-\delta) + h^4 (3-2\delta) \, .
$$
Si on inclut $\xi_0$ dans la perturbation, on doit ajouter sa contribution \`a $\frac{d}{dS} \, SDT$.

\smallskip

De $\vec\xi_0 = \frac{ih^2}{k} \, \frac{\vec U}{U^2} \, e^{iku}$ on tire :
$$
SD_0 \, T = - \, 2 \, k^2 \, V^2 \, \xi_u = + \, 2 \, V^2 \, h^2 \, e^{iku}
$$
$$
\frac{d}{dS} \, SD_0 \, T = -4 \, \ell^2 \, h^2 = - \, k^2 \, \delta^2 \, .
$$
Au total de $\xi_0$ et $\xi_1$,

\begin{center}
\begin{boxedminipage}{6cm} $\displaystyle \frac{d}{dS} \, SDT = - k^2 (2-\delta)^2 + h^4 (3-2\delta)$ \end{boxedminipage}
\end{center}

\medskip

On obtient la m\^eme valeur pour $\frac{d}{dS} \, {\rm div} \, \xi = + \frac{d}{dS} \, \frac{SDT}{V^2}$, du fait que le SDT total est d'ordre deux.

\newpage

\begin{center}
\bf Valeur r\'esiduelle de $D_m$
\end{center}

\medskip

Puisqu'on annule les \'el\'ements de $D$ qui ne sont pas purement m\'eridiens, seuls peuvent intervenir dans la valeur r\'esiduelle de $BD$ les \'el\'ements $D_{SS}$, $D_{tt}$ et $D_{St}$. Nous allons pr\'eciser leur contribution. Etant donn\'e qu'ils ont \'et\'e r\'eduits \`a l'ordre final, et qu'ils seront multipli\'es par $Q^*$ avant d'\^etre int\'egr\'es dans $\delta W$, ils peuvent \^etre d\'ecompos\'es harmoniquement sur l'angle polaire de la section, en ne retenant que l'harmonique homologue \`a l'harmonique dominante de $\xi$.

\smallskip

Si nous d\'efinissons, \`a l'aide du vecteur unit\'e normal $\vec e_n$ et du vecteur unit\'e tangent \`a la section $\vec e_t$ les vecteurs :
$$
\vec M = \vec e_n + i \, \vec e_t \qquad \hbox{et} \qquad \vec N = \vec e_n - i \, \vec e_t
$$
et si on appelle $\alpha$ l'angle $(\vec e_X , \vec e_n)$ on a :
$$
\vec M = \vec S \, e^{-i\alpha} \qquad \vec N = \vec T \, e^{i\alpha}
$$
d'o\`u l'on d\'eduit ais\'ement :
$$
4 \, i \, \vec e_t \, D \, \vec e_n = e^{-2i\alpha} \, SDS - e^{2i\alpha} \, TDT
$$
$$
-4 \, \vec e_t \, D \, \vec e_t = e^{-2i\alpha} \, SDS + e^{2i\alpha} \, TDT - 2 \, TDS \, .
$$
Les termes significatifs des r\'esidus de $SDS$, $TDT$, $SDT$ sont du 1$^{\rm er}$ degr\'e en $S$ et $t$ et varient donc angulairement comme $e^{-i\alpha}$ et $e^{i\alpha}$. D'autre part les termes qui contribuent au mode $m=1$ dans $B_m$ $D_m$ sont les termes en $e^{-i\alpha}$. On voit donc que $TDT$ ne contribue pas, que $SDS$ contribue par son terme en $t$ et $TDS$ par son terme en $S$.

\bigskip

\begin{center}
\bf La contribution de $\xi_0$
\end{center}

\medskip

On a not\'e plus haut que pour un champ $B_0 = U$, la composante $\xi \cdot U$ ne contribue pas \`a $G$. Autrement dit
$$
UD (\xi_0) = \nabla (\xi_0 \cdot U)
$$
[$\xi_0$ est-il parall\`ele \`a $U$ ou \`a $B$ ? (Ici \`a $U$, mais ne pourrait-on le prendre parall\`ele \`a $B$ ?)] Par un champ quelconque, on aura :
\begin{eqnarray}
G(\xi_0) &= &BD (\xi_0) - \nabla (\xi_0 \cdot B) \qquad (-(B\nabla)^{-1} \, {\rm div} \, \xi_0 \nabla p) \nonumber \\
&= &f \, U \cdot D + B_m \, D - \nabla (f \, \xi_0 \cdot U) \nonumber \\
&= &(f-f_0) \, U \cdot D + \nabla \, [(f_0 - f) \, \xi_0 \cdot U ] + B_m \cdot D \nonumber
\end{eqnarray}
o\`u $f_0$ est une constante arbitraire.

\smallskip

Par la r\'eciprocit\'e, on a $B_m \cdot D \cdot U = B_m \, \nabla (\xi_0 \cdot U)$, donc
$$
B_m \cdot D = \frac{\vec U}{U^2} \, B_m \, \nabla (\xi_0 \cdot U) + B_m \cdot D_m \, .
$$
On a vu que dans $D_m$, $\xi_0$ modifie seulement $SDT$, en le faisant passer \`a l'ordre deux et en changeant un peu les coefficients.

\smallskip

Les deux termes $(f-f_0) \, UD$ et $\frac{\vec U}{U^2} \, B_m \, \nabla (\xi_0 \cdot U)$, sont n\'egligeables : ils sont en effet d'ordre $\varepsilon^3 \, f \, \xi$ dans la direction m\'eridienne et $\varepsilon^2 \, f \, \xi$ dans la direction azimutale (\`a comparer \`a $\varepsilon^2 \, f \, \xi$ et $\varepsilon \, f \, \xi$ comme on le montre plus loin). On retiendra donc seulement :
$$
G(\xi_0) = \nabla \, [(f_0 -f) \, \xi_0 \cdot U] + B_m \, D_m (\xi_0) - (B\nabla)^{-1} \, {\rm div} \, \xi_0 \, \nabla p \, .
$$

\bigskip

\begin{center}
\bf Le calcul de $\delta W$
\end{center}

\medskip

Soit l'expression :
$$
\delta W = \frac{1}{2} \int d\tau \, Q^* \cdot G
$$
que nous appliquons \`a la fonction d'essai $\xi_0 + \xi_1$ dans le domaine limit\'e par la surface r\'esonante ($qk=1$) on a
$$
G = BD + \nabla [(f_0 - f) \, \xi_0 \cdot U] - \nabla (\xi_1 \cdot B) - (B\nabla)^{-1} \, {\rm div} \, \xi \, \nabla p \, .
$$
A l'ordre deux, $B \cdot D$ se r\'eduit \`a $B_m \, D_m (\xi_0 + \xi_1)$. Nous partageons l'int\'egrant en deux parties : 
$$
G = G_1 + G_2
$$
avec
$$
G_1 = B_m \, D_m - (p-p_1) \, \vec\nabla \, \eta \qquad B \nabla \eta + {\rm div} \, \xi = 0
$$
$$
G_2 = - \nabla (\xi_1 \cdot B) + \nabla [\eta (p-p_1)] - \nabla [(f-f_1) \, \xi_0 \cdot U]
$$
$p_1$ : valeur de la pression \`a la limite.

\smallskip

[On a remplac\'e $f_0$ par $f_1$, valeur de $f$ sur la surface r\'esonante.] Nous allons montrer que les termes de $G_1$ sont d'ordre deux. $G_2$ ne comporte que des gradients exacts qui donnent une int\'egrale de surface par int\'egration par parties. Indiquons d'abord que l'ordre significatif pour la composante m\'eridienne de $BD$, $(B \cdot D)_m$, est l'ordre deux, $(\varepsilon^2 \, f \, \xi)$ qui se combine avec l'ordre z\'ero $(f\xi)$ de $Q_m^*$. Pour la composante azimutale de $BD$, soit $(BD)_u$, c'est l'ordre un $(\varepsilon \, f \, \xi)$ qui est significatif, parce que $Q_u^*$ est d'ordre $\varepsilon \, f \, \xi$ (comme on le voit dans la m\'ethode altern\'ee).

\smallskip

Donc les termes d'ordre sup\'erieur \`a ceux-l\`a sont n\'egligeables (comme on l'a dit pour $\xi_0$).

\smallskip

Dans $G_1$, $B_m \, D_m$ est bien d'ordre $\varepsilon^2 \, f \, \xi$. Quant au second terme, il faut noter que ${\rm div} \, \xi$ est de l'ordre de $\xi$ et en outre constant \`a l'ordre principal, donc $\eta$ (en $\frac{\xi}{f}$) est aussi constant \`a l'ordre principal et $\nabla \eta$ reste de l'ordre de $\frac{\xi}{f}$ ; comme $p$ est en $\varepsilon^2 f^2$, on a bien le m\^eme ordre final. 

\smallskip

On int\`egre par parties le terme contenant $G_2$ et on obtient :
$$
\delta W = \frac{1}{2} \int d\tau \, Q^* \cdot G_1 + \frac{1}{2} \int dS \, \xi^* \cdot n \, B \nabla \, [ \xi_1 \cdot B] \, .
$$
On va calculer s\'epar\'ement le terme de volume et le terme de surface.

\bigskip

\begin{center}
\bf Evaluation des termes de volume
\end{center}

\medskip

Soit 
$$
\delta W = \frac{1}{2} \int d\tau \, Q_m^* \, D_m \, B_m - \frac{1}{2} \int d\tau \, Q_m^* \cdot \nabla \eta \, (p-p_1) \, .
$$
A l'ordre principal o\`u $\xi_m^*$ est constant, on a :
$$
Q_m^* \sim f \, \vec U \, \nabla \, \xi_m^* - \xi_m^* \, \nabla \, \vec B_m
$$
avec
$$
\xi_m^* \sim (\vec e_{\rho} + i \, \vec e_{\theta}) \, e^{i\theta - iku} \, \vert \xi_r \vert = \vert \xi_r \vert \, \vec S \, e^{-iku}
$$
[$B_m = S,T$, $\xi_m = S,T$, $Q_m = S,T$, $Q_m \, B_m \, D_m =$], ce qui donne (par unit\'e de longueur), pour le premier terme:
$$
\delta W_1 = \pi \, \vert \xi_r \vert^2 \int_0^{\rho_1} \rho \, d\rho \, \vert B_m \vert \left\{ - \frac{d \, \vert B_m \vert}{d\rho} \, \vec e_{\theta} + i \, \frac{\vert B_m \vert}{\rho} \, \vec e_{\rho} - ikf (\vec e_{\rho} + i \, \vec e_{\theta}) \right\} \cdot [D_m \cdot \vec e_{\theta} \, e^{i\theta}] \, .
$$
En substituant $\vert B_m \vert = \frac{\rho f}{q}$ et $\langle D_m \cdot \vec e_{\theta} \, e^{i\theta} \rangle = - \frac{f}{2} \, k^2 \, [i \, \alpha \, \vec e_{\rho} + (\alpha + 2\beta) \, \vec e_{\theta}]$ on obtient, apr\`es une int\'egration par parties sur $q$ :
$$
-\delta W_1 = \pi \, \vert \xi_r \vert^2 \, f^2 \, k^4 \left\{ \alpha \int_0^{\rho_1} \rho^3 \, d\rho \left( \frac{1}{kq} - 1 \right) + \beta \int_0^{\rho_1} \rho^3 \, d\rho \left( \frac{1}{kq} - 1 \right) \left( \frac{1}{kq} + 2 \right) \right\} \, .
$$
Pour le terme de pression, on a :
\begin{eqnarray}
-\delta W_2 &= &\frac{1}{2} \int d\tau \, Q^* \cdot \nabla \eta \, (p-p_1) \nonumber \\
&= &\frac{1}{2} \int d\tau \, (p-p_1) \, {\rm div} \, (\xi^* \, \nabla \eta \, \vec B - \vec\xi^* B \, \nabla \eta) \nonumber \\
&= &\frac{1}{2} \int d\tau \, (p-p_1) \, {\rm div} \, (\vec\xi^* \, {\rm div} \, \xi) \nonumber \\
&= &\frac{1}{2} \int d\tau \, (p-p_1) \, [{\rm div} \, \xi_m^* \, {\rm div} \, \xi + \xi_m^* \nabla \, {\rm div} \, \xi] \, . \nonumber
\end{eqnarray}
Or on a \'etabli
$$
{\rm div} \, \xi = - \frac{S \cdot D \cdot T}{V^2} - \frac{U \cdot D \cdot U}{U^2} \sim + \, h^2 \, e^{iku}
$$
et puisque ${\rm div} \, \xi_u \sim {\rm div} \, \xi_0 = -h^2 \, e^{iku}$, ${\rm div} \, \xi_m \sim 2 \, h^2 \, e^{iku}$. Enfin
$$
\xi_m^* \, \nabla \, {\rm div} \, \xi \sim \frac{\vec S}{V^2} \, \nabla \left( - \frac{S \cdot D \cdot T}{V^2} - \frac{U \cdot D \cdot U}{U^2} \right) = 4 \, k^2 \beta \, .
$$
Au total, on a :
$$
-\delta W^2 = \xi_r^2 \left( \frac{h^4}{2k^2} + \beta \right) \frac{1}{2} \int d\tau (p-p_1) \, 4 \, k^2 \, .
$$

\bigskip

\begin{center}
\bf Calcul du terme de surface
\end{center}

\medskip

Ce terme s'\'ecrit :
$$
\delta W_{S_1} = \frac{1}{2} \int dS \, \xi_1^* \cdot n \, B \nabla (\xi_1 \cdot B) \, .
$$
Cette int\'egrale doit \^etre effectu\'ee sur la surface magn\'etique exacte o\`u le mode $m=1$ est r\'esonant ($kq=1$).

\smallskip

L'op\'erateur $B\nabla$ fait cro{\^\i}tre d'un ordre le second facteur, puisque l'harmonique $m=1$ est dominante dans $\xi_1$. [D'accord mais avec quelle fonction-test ?] Et de ce fait, l'int\'egrale prend automatiquement une valeur d'ordre deux. [Laval : $\psi$ et $\frac{d\psi}{dn}$ ou $\frac{d\psi}{dF}$.] Pour tirer parti correctement de cette propri\'et\'e, il suffit d'effectuer cette partie du calcul en coordonn\'ees intrins\`eques de l'\'equilibre. Plus pr\'ecis\'ement, on d\'ecompose harmoniquement en $\chi$ chacun des deux facteurs (\`a l'ordre un). Dans le syst\`eme $\mu , u , \chi$ on a :
$$
B\nabla = {\mathcal J} \left( ik + \frac{1}{q} \, \frac{\partial}{\partial\chi} \right) \qquad dS = \frac{d\tau \, \vert \nabla \mu \vert}{d\mu} = \frac{f\mu \, \vert \nabla \mu \vert}{\mathcal J} \, du \, d\chi
$$
$$
\delta W_{S_1} = \frac{f}{2} \int du \, d\chi \, (\mu \, \xi_1^* \, \nabla \mu) \left( ik + \frac{1}{q} \, \frac{\partial}{\partial\chi} \right) (\xi_1 \cdot B) \, .
$$
Par unit\'e de longueur en $u$ (avec $kq = 1$)
$$
\delta W_{S_1} = \frac{fk}{2} \int d\chi \, [\mu \, \xi_1^* \, \nabla \mu] \left[ i \, \frac{d}{\partial\chi} - 1 \right] [-i \, \xi_1 \cdot B]
$$
[$\xi \cdot B$ et $\xi \cdot \nabla F$ sont les 2 quantit\'es continues (\`a $B \nabla$ pr\`es) sur la r\'esonance.] L'harmonique $m=1$ ne contribue pas. A l'ordre consid\'er\'e, interviennent seulement $m=0$ et $m=2$. L'harmonique $m=0$ ne joue pas (il n'y a pas de mode de compression par construction, et le mode de rotation a \'et\'e \'elimin\'e par $\xi_0$).

\smallskip

Il reste donc seulement \`a calculer l'apport du mode $m=2$. On \'ecrira pour cela $\mu \, \xi_1^* \cdot \nabla \mu$ et $i \, \xi_1 \cdot B$ d'abord dans les coordonn\'ees $s,u,t$, puis on effectuera le changement de coordonn\'ees en $\mu , u , \chi$. On pourra d'ailleurs faire d'embl\'ee $t=0$, puisque seuls les termes en $1,S,S^2$ contribueront significativement \`a $e^{-2i\chi}$.

\smallskip

On peut \'ecrire ({\it cf.} Annexe C) :
$$
i \, \xi_m = \nabla W \times \frac{\vec U}{U^2} \ \Biggl\vert \ X = \mu \, \xi \cdot \nabla \mu = i \, \frac{\partial W}{\partial \chi}
$$
$$
i \, \xi \cdot B = i \, \xi \cdot U f + i \, \xi \cdot B_m = \frac{f}{q} \left[ \left( 1 + i \, \frac{\partial}{\partial \chi} \right) \xi_u + \frac{2h\ell}{k} \, U^2 \, i \, \frac{\partial W}{\partial \chi} \right]
$$
(expressions valables aux deux premiers ordres ; $W$ et $\xi_u$ n'ont pas de composante en $m=0$. Par contre $i \xi \cdot B$ en poss\`ede une d'ordre deux, qu'il faut supposer absorb\'ee dans $\xi_0$). Pour d\'eterminer $W(S)$ on utilise
$$
\xi_S = \frac{V^2}{U} \, \frac{\partial W}{\partial S} - kKW \qquad W(S) = 2S + (2\ell^2 - h^2) \, S^2 \, .
$$
D'autre part :
\begin{eqnarray}
\xi_u &= &2 \left( 1 - \frac{2\ell h}{k} \right) S + 2 \, S^2 \left[ \ell^2 - h^2 - \frac{\ell h}{k} \, (2\ell^2 - 3h^2) \right] \nonumber \\
U^2 &= &1 - 2 \, h^2 S \nonumber \\
2S &= &\mu \, e^{-i\chi} + \frac{\mu^2}{2} \, e^{-2i\chi} (h^2 \lambda + 2h^2 - \ell^2) + \cdots \nonumber
\end{eqnarray}
ce qui donne par substitution :
\begin{eqnarray}
W &= &\mu \, e^{-i\chi} + \frac{\mu^2 k^2}{2} \, e^{-2i\chi} \left( \lambda + \frac{3}{2} \right) \nonumber \\
\xi_u &= &\mu \, e^{-i\chi} \left( 1 - \frac{2h\ell}{k} \right) + \frac{\mu^2 h^2}{2} \, e^{-2i\chi} \left[ \lambda + 1 - \frac{2h\ell}{k} \, \left( \lambda + \frac{1}{2} \right) \right] \nonumber \\
X^* &= &\mu \, \xi^* \, \nabla \mu = \mu^2 \, h^2 \, e^{2i\chi} \left( \lambda + \frac{3}{2} \right) \qquad \hbox{pour} \ m=2 \nonumber \\
+ \, i \, \phi &= &-i \, \xi \cdot B = -fk \, \mu^2 \, h^2 \, e^{-2i\chi} \left[ \frac{3}{2} \, (\lambda + 1) - \frac{h\ell}{k} \left( \lambda + \frac{1}{2} \right) \right] \quad \hbox{pour} \ m=2 \nonumber \\
\delta \, W_{S_1} &= &-\pi \, f^2 k^2 \mu^4 h^4 \left[ \frac{3}{2} \, (\lambda + 1) - \frac{h\ell}{k} \left( \lambda + \frac{1}{2} \right) \right] \left( \lambda + \frac{3}{2} \right) \, . \nonumber
\end{eqnarray}

\bigskip

\begin{center}
\bf Compl\'ement de la solution
\end{center}

\medskip

La solution compl\`ete doit satisfaire les conditions aux limites, ainsi que les conditions de continuit\'e sur la surface r\'esonante : continuit\'e de $Q \cdot \nabla F$ et de $B \cdot G$ (pression totale perturb\'ee), autrement dit de $B \cdot \nabla \phi$ pour une solution de l'\'equation d'Euler. Il est donc n\'ecessaire d'ajouter \`a $\xi_0 + \xi_1$ une troisi\`eme partie $\xi_2$, qui d\'ependra de l'\'equilibre. Si on d\'ecompose la solution en harmoniques de l'angle intrins\`eque $\chi$, l'harmonique dominante $m=1$ n'a pas de condition de continuit\'e \`a satisfaire et sera nulle au-del\`a de la surface r\'esonante en raison des conditions aux limites. Pour l'harmonique $m=0$, $Q \cdot \nabla F$ et de $B \cdot G$ sont nuls sur cette surface, donc aussi au-del\`a pour la m\^eme raison que $m=1$. Ainsi $\vec\xi_2$ ne contient qu'une harmonique $m=2$, de part et d'autre de la surface. Son amplitude est d'ordre $\varepsilon$ devant $\vec \xi_1$. Il suffit par suite de la calculer \`a l'ordre principal m\'eridien, suivant la m\'ethode de l'Annexe C. La contribution de $\xi_2$ \`a $\delta W$ appara{\^\i}t comme un terme de surface compl\'ementaire. On a en effet (en n'\'ecrivant pas $\xi_0$)
$$
\delta W = \delta W (\xi_1) + 2 \, \delta W (\xi_1 , \xi_2) + \delta W (\xi_2)
$$
avec
\begin{eqnarray}
\delta W (\xi_1 , \xi_2) &= &\frac{1}{2} \int dS \, \xi_{2n}^* \, B \nabla (\xi_1 \cdot B) \nonumber \\
\delta W (\xi_2) &= &\frac{1}{2} \int dS \, \xi_{2n}^* (- B \nabla \phi_2) \nonumber
\end{eqnarray}
puisque le terme de volume du couplage est d'ordre sup\'erieur (\`a condition d'\'ecrire l'int\'egrant dans le sens convenable). On doit d\'eterminer pour $\xi_2$ la solution int\'erieure $\xi_i$ et la solution ext\'erieure $\xi_e$. Elles sont fix\'ees par les conditions \`a remplir sur la r\'esonance (ainsi que les conditions aux limites et les conditions de passage sur une \'eventuelle surface r\'esonante $m=2$). On doit avoir sur la r\'esonance :
$$
X_i + X_1 = X_e \qquad\qquad \tilde\phi_i + \tilde\phi_1 = \tilde\phi_e
$$
($X = \mu \, \xi \, \nabla\mu$, $\tilde\phi = i \, \phi$) ($X_1$ et $\tilde\phi_1$ \'etant les composantes $m=2$ dans $\vec\xi_1$). Les conditions sur l'axe et sur la fronti\`ere d\'eterminent les relations $\tilde\phi_i = A \, X_i$ et $\tilde\phi_e = B \, X_e$ ($A$ et $B$ d\'ependent du profil de courant). En calculant et en ajoutant les diff\'erents termes de surface, on montre ais\'ement qu'on obtient le total en rempla\c cant dans l'int\'egrale de surface de $\xi_1$ seul le produit $\xi_1 \, \tilde\phi_1$ par $\xi_e \, \tilde\phi_1 - \xi_1 \, \tilde\phi_e$, c'est-\`a-dire :
$$
\frac{[\tilde\phi_1 - A \, X_1] \, [\tilde\phi_1 - B \, X_1]}{B-A}
$$
ce qui donne :
$$
\delta W_S = \pi k f \, \frac{1}{B-A} \, [\tilde\phi_1 - A \, X_1] \, [\tilde\phi_1 - B \, X_1]
$$
avec
\begin{eqnarray}
- \, \tilde\phi_1 &= &fk \, \mu^2 h^2 \left[ \frac{3}{2} \, (\lambda + 1) - \frac{h\ell}{k} \left( \lambda + \frac{1}{2} \right)\right] \nonumber \\
X_1 &= &\mu^2 h^2 (\lambda + 3/2) \nonumber
\end{eqnarray}
$A$ et $B$ sont calcul\'es dans l'Annexe C. Pour $m=2$ et $kq = 1$, on montre que :
$$
A \ {\rm ou} \ B = kf \left[ \frac{1}{4} \, \frac{\mu}{\xi \nabla \mu} \, \frac{d}{d\mu} \, \xi \, \nabla \mu + \frac{h\ell}{k} - \frac{3}{4} \right] \, .
$$
Pour un courant plat on a :
$$
\left[ \frac{1}{4} \, \frac{\mu}{\xi \nabla \mu} \, \frac{d}{d\mu} \, \xi \, \nabla \mu \right]_i = \frac{1}{4} \qquad \left[ \frac{1}{4} \, \frac{\mu}{\xi \nabla \mu} \, \frac{d}{d\mu} \, \xi \, \nabla \mu \right]_e = -\frac{3}{4} 
$$
on posera donc
\begin{eqnarray}
A &= &kf \left( \frac{h\ell}{k} - \frac{1}{2} + a \right) \nonumber \\
B &= &kf \left( \frac{h\ell}{k} - \frac{3}{2} + b \right) \, . \nonumber
\end{eqnarray}
En rempla\c cant d'autre part $\lambda$ par $\tilde\lambda - \frac{3}{4} - \frac{h\ell}{k}$, on obtient :
$$
\delta W_S = + \, \pi \, k^4 f^2 \mu^4 \, \xi_r^2 \, \frac{h^4}{k^2} \, \frac{\left[ \tilde\lambda + a \left( \tilde\lambda + \frac{3}{4} - \frac{h\ell}{k} \right)\right] \left[\frac{3}{4} - \frac{h\ell}{k} + b \left( \tilde\lambda + \frac{3}{4} - \frac{h\ell}{k} \right)\right]}{1+a-b} \, .
$$

\bigskip

\begin{center}
\bf Regroupement des r\'esultats
\end{center}

\medskip

Pour condenser les formules nous mettons \`a part le facteur de normalisation de $\delta W$ qui est \'egal par unit\'e de $u$ \`a $\pi f^2 k^4 \mu^4 \, \xi_r^2$. Et nous posons

\medskip

\begin{center}
\begin{boxedminipage}{15mm} $\displaystyle \frac{h\ell}{k} = \delta$ \end{boxedminipage} \, .
\end{center}

\medskip

\noindent Nous avons obtenu $\delta W$ comme une somme de 3 termes :

\smallskip

\noindent 1) un terme de courant : $\delta W_1 = \alpha \, \frac{S}{4} + \beta \left( \frac{S}{4} + T \right)$ avec
\begin{eqnarray}
S &= &4 \int_0^1 z^3 \, dz \left( \frac{1}{kq} - 1 \right) \nonumber \\
T &= &\int_0^1 z^3 \, dz \left( \frac{1}{k^2 q^2} - 1 \right) \nonumber \\
\alpha &= &-(1-\delta)^2 \, (1-4\delta) + \frac{h^4}{k^2} \left( \frac{5}{4} - 4 \delta + 4 \delta^2 \right) \nonumber \\
\beta &= &-(1-\delta)^2 + \frac{h^4}{k^2} \left( \frac{3}{4} - \delta \right) \nonumber
\end{eqnarray}

\noindent 2) un terme de pression : $\delta W_2 = Y^P$ avec
\begin{eqnarray}
P &= &4 \int_0^1 z \, dz \, \frac{p-p_1}{B_{m_1}^2} \nonumber \\
Y &= &\beta - \frac{h^4}{2k^2} = -(1-\delta)^2 + \frac{h^4}{k^2} \left( \frac{1}{4} - \delta \right) \nonumber
\end{eqnarray}

\noindent 3) un terme de surface
$$
\delta W_3 = \frac{1}{1+a+b} \, \frac{h^4}{k^2} \left\{ \tilde\lambda + a \left( \tilde\lambda + \frac{3}{4} - \delta \right)\right\} \left\{ \frac{3}{4} - \delta + b \left( \tilde\lambda + \frac{3}{4} - \delta \right)\right\}
$$
avec $\tilde\lambda = T - \delta S + P$.

\smallskip

Cet ensemble se met sous la forme :
$$
\frac{k^2}{h^4} \, \delta W = \left[ \tilde\lambda + \frac{S}{2} \right] \left[ (1-2\delta) - (1-\delta)^2 \, \frac{k^2}{h^4} \right] + \frac{T}{2} + \frac{1}{1+a-b} \left\{ a \left( \frac{3}{4} - \delta \right)^2 - b \, \tilde\lambda^2 - ab \left( \tilde\lambda + \frac{3}{4} - \delta \right)^2 \right\} \, .
$$

\begin{enumerate}
\item[--] Si on fait $h=0$, $(\delta = 0)$, on retrouve le cylindre :
$$
\delta W = - \left(\tilde\lambda + \frac{S}{2} \right) = - \left( T+P+ \frac{S}{2} \right) = - \int_0^1 dz \, z^3 \left( \frac{1}{k^2 q^2} + \frac{2}{kq} - 3 \right) - 4 \int z \, dz \, \frac{p-p_1}{B_m^2} \, .
$$
\item[--] Si on fait $\frac{h^4}{k^2} = 0$ mais $\delta \ne 0$ (h\'elice infinie, Stellarator) on obtient :
$$
\delta W = -(1-\delta)^2 \left[ T+P+\left( \frac{1}{2} - \delta \right) S \right] \ \Biggl\vert \ \hbox{stable pour} \ S > \frac{(T+P)}{S} + \frac{1}{2}
$$
effet stabilisant pour $\delta > 0$.
\item[--] Si on fait $\ell = 0$, on retrouve le tore : ($\delta = 0$)
$$
k^2 \, \delta W = \left[ T+P+\frac{S}{2} \right] [1-k^2] + \frac{T}{2} + \frac{\frac{9}{16} \, a - b \, (T+P)^2 -ab \left( T+P+ \frac{3}{4} \right)^2}{1-a-b} \, .
$$
\item[--] Si on fait $\delta = 1$ (\'equilibre sans courant)
\end{enumerate}

\bigskip

\begin{center}
\bf Discussion -- Interpr\'etation
\end{center}

\medskip

La formule contient deux sortes de param\`etres :

\smallskip

\noindent 1) des param\`etres caract\'eristiques du profil de courant (et de pression). Ce sont les m\^eme qu'en g\'eom\'etrie torique, \`a savoir : $S,T,P,a,b$

\smallskip

\noindent 2) deux param\`etres de forme, un pour l'h\'elicit\'e $\frac{h}{\ell}$, l'autre pour la p\'eriode normalis\'ee de la perturbation, $\delta = \frac{h\ell}{k}$ $\left( \frac{h^4}{k^2} = \left( \frac{h}{\ell} \, \delta \right)^2 \right)$, ou p\'eriode des lignes magn\'etiques r\'esonantes, rapport\'ee \`a la p\'eriode $hk$ des lignes magn\'etiques d'un {\it champ sans courant} (lorsque le courant axial total d\^u au champ polo{\"\i}dal compense le courant d\^u \`a $B = \vec U$). Ce champ correspond \`a $\delta = 1$ et constitue en fait le v\'eritable champ de r\'ef\'erence (le champ en l'absence de plasma).

\smallskip

D'ailleurs on constate que pour $\delta = 1$ (\'equilibre sans courant), $\delta W$ est (\`a un facteur pr\`es) ind\'ependant de l'h\'elicit\'e.

\medskip

\noindent 1) Pour discuter le r\'esultat obtenu, nous observerons d'abord l'effet d\'estabilisant de la pression (\`a travers le terme $b \, \tilde\lambda^2$ essentiellement) et nous examinerons ensuite ce qui se passe pour $P=0$. 

\smallskip

\noindent 2) Nous supposerons le shear assez faible pour que $a$ et $b$ soient petits, ainsi que $T$ et $S$. Nous retiendrons les termes du premier ordre pour obtenir :
$$
\frac{\delta W}{4S} = - (1-\delta)^3 + \frac{3h^4}{k^2} \left( \delta^2 - \frac{3}{2} \, \delta + \frac{29}{48} \right)
$$
o\`u on a fait $T \sim 2S$ et $a \sim 4S$ (shear faible).

\smallskip

On v\'erifie que le coefficient de $\frac{h^4}{k^2}$ est toujours positif. On pose donc $\frac{h^4}{k^2} = \delta^2 \, \frac{h^2}{\ell^2}$ et on a la condition de stabilit\'e :
$$
\frac{h^2}{\ell^2} > \frac{(1-\delta)^3}{3 \, \delta^2 \left\{ \delta^2 - \frac{3}{2} \, \delta + \frac{29}{48} \right\}} = F(\delta)
$$
d'o\`u l'on d\'eduit que la condition est toujours remplie pour $\varepsilon > 1$ (h\'elicit\'e des lignes magn\'etiques invers\'ee par rapport \`a $U$ et {\it courant principal invers\'e}).

\smallskip

Dans les autres cas, on peut s'assurer que la fonction $F(\delta)$ d\'ecro{\^\i}t de $+\infty$ \`a $0$ quand $\delta$ varie de z\'ero \`a un, et d\'ecro{\^\i}t de $+ \infty$ \`a z\'ero quand $\delta$ va de z\'ero \`a $-\infty$ (h\'elicit\'e de m\^eme signe). Il en r\'esulte qu'une valeur donn\'ee de $\frac{h^2}{\ell^2}$ stabilise tous les $\delta$ assez grands (en module) c'est-\`a-dire les modes de p\'eriode assez grande. Et la stabilisation est d'autant meilleure que $\frac{h^2}{\ell^2}$ est \'elev\'e.

\smallskip

Si on compare maintenant avec le cas de la g\'eom\'etrie torique (quasi-marginale), on remarque que la stabilisation provient ici de nouveaux effets, \`a savoir essentiellement :

\smallskip

\noindent 1) du fait que la courbure $\left( \frac{1}{n^2} \right)$ est remplac\'ee par $\frac{h^4}{k^2}$ qui peut \^etre grand en pratique, m\^eme pour une h\'elicit\'e r\'eduite parce que le pas de l'h\'elice peut \^etre notablement plus court que le pas de la perturbation r\'esonante (des lignes magn\'etiques), tandis que dans un tore ils sont n\'ecessairement identiques. [Torsion?]

\smallskip

\noindent 2) du r\^ole de $\delta$ dans le formule, en particulier la stabilisation pour $\delta >1$.

\smallskip

Par exemple, pour $\delta \to -\infty$ on a la condition :
$$
- \frac{1}{\delta} \, \frac{\ell^2}{h^2} < 2 + \frac{\frac{a}{4S} + 4 \, b \, S + \frac{ab}{4S} \, (4S+ 1)^2}{1+a+b} \, .
$$
[Si on pose $Q = 4S\delta - (2S+T+P)$ et $\delta = 1+\eta$ \c ca fait :
$$
\delta W \, \frac{k^2}{h^4} = Q \left[ \eta^2 \, \frac{k^2}{h^4} + 1 + 2\eta \right] + \frac{T}{2} + (1+a+b)^{-1} \left\{ a \left( \eta + \frac{1}{4} \right)^2 + b \, (Q+2S)^2 + ab \left( Q + 2S + \eta + \frac{1}{4} \right)^2 \right\} \, .
$$
Du 2$^{\rm e}$ degr\'e en $Q$ ou en $\frac{Q}{S} $.]

\smallskip

Lorsque $\delta > 1 + \frac{P+T-2S}{4S}$ le terme ind\'ependant de $\frac{h^2}{\ell^2}$ est positif. Alors on est stable si $\frac{h^4}{k^2}$ n'est pas trop grand, ou si son facteur est positif aussi, c'est-\`a-dire essentiellement $b \, \tilde\lambda^2$ pas trop grand (on peut \'ecrire la condition $\vert b \, \lambda^2 \vert <$).

\bigskip

\begin{center}
\bf Annexe A -- Coordonn\'ees h\'elico{\"\i}dales
\end{center}

\medskip

Posons
$$
\frac{\nabla u \cdot \nabla v}{(\nabla v)^2} = \frac{h\ell (r^2 - 1)}{\ell^2 + h^2 r^2} = K(r) \qquad \left( \frac{dK}{dr^2} = h\ell \, U^4 \right) \, .
$$
Soit
$$
\vec X = \vec{\nabla} x \qquad \vec U = \vec{\nabla} u - K \vec{\nabla} v = \frac{\ell \, \vec e_z + hr \, \vec e_{\varphi}}{\ell^2 + h^2 r^2} \qquad \vec V = \vec{\nabla} v = h \, \vec e_z - \frac{\ell}{r} \, \vec e_{\varphi} \, .
$$
Soit $U = \vert \vec U \vert$ et $V = \vert \vec V \vert = \frac{dx}{dr} = \vert \vec X \vert$, $\vec X$, $\vec U$, $\vec V$ sont orthogonaux (mais non norm\'es). ($U^2 = \frac{1}{\ell^2 + h^2 r^2}$, $rUV = 1$). [Commutateurs : $(U,V) = 0$, $\left( U , \frac{R}{r} \right) = \nabla V^2 \times V$, $\left( \frac{R}{r} , V \right) = {\rm rot} \, (V^2 \, \vec U)$.]

\smallskip

\noindent On a ${\rm div} \, \vec U = 0$, ${\rm div} \, (U \, \vec X) = 0$, ${\rm div} \, \vec V = 0$, ${\rm rot} \, \vec X = 0$, ${\rm rot} \, \vec V = 0$, ${\rm rot} \, \vec U = 2 h \ell \, U^2 \, \vec U$. 

\smallskip

\noindent On utilisera : $\frac{dU}{dx} = -h^2 \, \frac{U^2}{V^2}$, $\frac{dV}{dx} = -\ell^2 \, U^3 \, V$, $K = h\ell \, U^2 (r^2 - 1)$, $\frac{dK}{dx} = -2h\ell \, \frac{U^3}{V^2}$, $\nabla S = \frac{T}{2}$, $\nabla t = \frac{S}{2}$. Avec $S = \frac{x-iV}{2}$, $t = \frac{x+iV}{2}$ on a : 
$$
x = S+t \qquad V = i(S-t) \qquad \frac{d}{dS} = \frac{d}{dx} + i \, \frac{d}{dV} \qquad \frac{d}{dt} = \frac{d}{dx} - i \, \frac{d}{dV} \, .
$$
Soit $\vec S = \vec X + i \, \vec V$, $\vec U$, $\vec T = \vec X - i \, \vec V$ la base complexe. Toutes les quantit\'es suivantes sont nulles :
$$
\vec S^2 \, , \ \vec T^2 \, , \ \vec S \cdot \vec U \, , \ \vec T \cdot \vec U \, , \ {\rm rot} \, \vec S \, , \ {\rm rot} \, \vec T \, , \ {\rm div} (U \vec S) \, , \ {\rm div} (U \vec T) \, , \ {\rm div} \, \vec U \, .
$$
On a : $\vec S \cdot \vec T = 2 \, V^2$, plus $\vert \vec U \vert^2 = U^2$, ${\rm rot} \, \vec U = 2h\ell \, U^2 \, \vec U$. [$S \nabla S = 0$, $T \nabla T = 0$.] On a encore $\vec S \times \vec U = -i \, U \vec S$ et $\vec T \times U = i \, U \, \vec T$
$$
\vec S \, \nabla f = V^2 \left[ \frac{\partial f}{\partial S} + i \, K \, \frac{\partial f}{\partial u} \right] \, \vec U \, \nabla f = U^2 \, \frac{\partial f}{\partial u}
$$
$$
\vec T \, \nabla f = V^2 \left[ \frac{\partial f}{\partial t} - i \, K \, \frac{\partial f}{\partial u} \right] \, S \times T = 2i \, \frac{V^2}{U} \, \vec U \, .
$$
Pour les quantit\'es de l'\'equilibre $\frac{\partial f}{\partial u} = 0$, et pour les perturbations ($\vec\xi = e^{iku} \, \vec\xi (r,V)$), $\frac{\partial f}{\partial u} = ikf$ et $\vec U \, \nabla f = i \, k \, U^2 f$. (Donner le tableau $A \nabla B$ pour la base $R,U,V$.)

\bigskip

\begin{center}
\bf Annexe B -- Equilibres \`a ``cercles d\'ecentr\'es''
\end{center}

\medskip

Comme en g\'eom\'etrie  torique, les \'equilibres \`a fronti\`ere circulaire ont des surfaces magn\'etiques dont les sections m\'eridiennes sont des cercles d\'ecentr\'es, aux deux premiers ordres (du moins formellement, dans les coordonn\'ees $x$ et $V$ qui ne sont pas de vraies distances). [C'est vrai aussi pour un cercle r\'eel, car l'ellipticit\'e est petite.] A l'ordre principal, le flux $F=F_0$ satisfait :
$$
\frac{d^2 F_0}{dx^2} + \frac{d^2 F_0}{dV^2} + 2h\ell \, f(F_0) + f \, \frac{df}{dF} \, (F_0) + \frac{dp}{dF} \, (F_0) = 0 \, .
$$
Posons $x = \rho \cos \theta$, $V = \rho \sin \theta$.

\smallskip

\noindent A cet ordre, les solutions \`a fronti\`ere circulaire ne d\'ependent que de $\rho$. Soit $F_0 (\rho)$ l'une d'elles. On a :
$$
\frac{d^2 F_0}{d\rho^2} + \frac{1}{\rho} \, \frac{d F_0}{d\rho} + 2 h \ell f (F_0) + f \, \frac{df}{dF} \, (F_0) + \frac{dp}{dF} \, (F_0) = 0 \, . \leqno (1)
$$
A l'ordre suivant, comme tous les termes correctifs qui constituent le second membre sont en $\cos \theta$, il en est de m\^eme pour la modification \`a apporter \`a $F_0$. Il est commode de chercher l'ensemble du flux modifi\'e sous la forme :
$$
F = F (\rho , \theta) = F_0 [\rho - \Delta (\rho) \cos \theta] = F_0 [\sigma (\rho , \theta)] \, .
$$
Le d\'ecentrement $\Delta (\rho)$, d'ordre $\varepsilon \rho$, du ``cercle'' $\sigma =$ cte (en r\'ealit\'e une ellipse) sera r\'egi par une \'equation qu'on obtient en \'ecrivant que $F$ satisfait l'\'equation d'\'equilibre aux deux premiers ordres. Par d\'erivation de l'expression de $F$, on a d'abord :
$$
\frac{1}{\rho} \, \frac{d}{d\rho} \, \rho \, \frac{dF}{d\rho} + \frac{1}{\rho^2} \, \frac{d^2 F}{d \theta^2} - \left( \frac{d^2 F_0}{d \sigma^2} + \frac{1}{\sigma} \, \frac{dF_0}{d \sigma} \right) + \cos \theta \left[ F'_0 \, \Delta'' + 2 \, F''_0 \Delta' + \frac{F'_0 \, \Delta'}{\rho} \right] \ne 0 (\varepsilon) \, .
$$
D'autre part, en prenant $\sigma$ pour argument de $F_0$ dans (1), on a
$$
\frac{d^2 F_0}{d\sigma^2} + \frac{1}{\sigma} \, \frac{d F_0}{d\sigma} + 2h\ell \, f(F_0) + f \, \frac{df}{dF} + \frac{dp}{dF} \, (F) = 0 \, .
$$
Nous additionnons ces deux \'egalit\'es et nous d\'eduisons de leur somme l'\'equation de l'\'equilibre exact (qui doit \^etre valable pour $F$ \`a deux ordres). Nous obtenons :
$$
\cos \theta \left[F'_0 \, \Delta'' + 2 \, F''_0 \Delta' + \frac{F'_0 \, \Delta'}{\rho} \right] + h^2 \, U \,  \frac{dF}{dx} + (1 - V^2) \, L_0 \, F + 2h\ell \, f (1-U^2) + \left( 1 - \frac{1}{U^2} \right) \frac{dp}{dF} \, .
$$
En retenant l'ordre principal (termes en $\cos \theta$), en multipliant par $\rho \, F'_0$ et en int\'egrant, il vient $(L_0 \, F \sim F'' + \frac{1}{\rho} \, F')$ :
$$
\rho_1 \, F'^2_1 \Delta'_1 + \ell^2 \, \rho_1^2 \, F'^2_1 + h^2 \int_0^{\rho_1} d\rho \left[ \rho \, F'^2 + 4h\ell \, \rho^2 fF' - 2\rho^2 \, \frac{dp}{d\rho} \right] = 0 \, .
$$
Nous posons $F' = -\rho \, \frac{f}{q}$ ($F'$ est n\'egatif) et $z = \frac{p}{\rho_1}$. Nous obtenons : $h\ell q_1 = \delta$, $\frac{q_1}{q} = Q$, $\frac{dp}{dz} \, \frac{1}{F'^2_1} = P'$
$$
- \frac{\Delta'_1}{\rho_1} = \ell^2 + h^2 \int_0^1 z^3 \, dz \left\{ Q^2 - 4 \delta \, Q - 2 \, \frac{1}{z} \, P' \right\} \, .
$$
La modulation du champ polo{\"\i}dal sur chaque surface magn\'etique est li\'ee directement au d\'ecentrement. $\vec B_p$ \'etant de la forme $\vec U \times \vec\nabla F (\sigma)$ avec $\sigma = \rho - \Delta \cos \theta$, la modulation relative de $\vert B_p \vert$ est la somme de celle de $\vert U \vert$ et de celle de $\vert \nabla F \vert$ (\`a $\sigma$ constant), c'est-\`a-dire de $\nabla \sigma$, ou de $\nabla \rho \, (1 - \Delta' \cos \theta)$ (car la contribution de $\nabla \theta$ n'intervient qu'au second ordre). Or $\vert \nabla \rho \vert$ est modul\'e comme $\vert \nabla x \vert$ et $\vert \nabla V \vert$ c'est-\`a-dire comme $V$ (l'ellipticit\'e des ``cercles''!), de sorte qu'au total $\vert B_p \vert$ est modul\'e comme $UV (1-\Delta' \cos \theta) \sim 1 - (\rho + \Delta') \cos \theta$. Nous \'ecrivons cette expression sous la forme $1+h^2 \lambda \rho \cos \theta$ en posant $\Delta' + \rho \, (1 + \lambda \, h^2) = 0$. Soit
$$
\lambda + 1 = \int_0^1 z^3 \, dz \, \{ Q^2 - 4\delta \, Q \} - \int_0^1 2z^2 \, dz \, [P'] \, .
$$
Cette quantit\'e interviendra dans la deuxi\`eme partie. Les \'equilibres obtenus sont valables lorsque la perturbation de $F$ reste petite devant $F_0$ on constate que la modulation du champ polo{\"\i}dal est en $h^2$.

$\cdots$

 On aboutit \`a la m\^eme conclusion par la minimisation de $\delta W$. En effet la contribution azimutale positive $Q_u^* \cdot G_u$ doit \^etre r\'eduite \`a l'ordre $\varepsilon^4$, \'etant donn\'e l'ordre du couplage de $Q_u^*$ et $G_u$ avec $Q_m^*$ et $G_m$.

\smallskip

En g\'eom\'etrie h\'elico{\"\i}dale, (2) s'\'ecrit :
$$
U \cdot G = B_m \nabla (\xi \cdot U) - f \, {\rm div} \, [\xi_m \, U^2] + 2h\ell \, U^4 \, \xi \cdot \nabla F = 0 (\varepsilon^2) \, .
$$
Puisqu'il y a un arbitraire dans $\vec\xi$, on pourra supposer que $\xi \cdot U$ est d'ordre $\varepsilon \, \xi_m$, ou alors d'ordre $\xi_m$ mais constant \`a cet ordre. D\`es lors tous les termes autres que la divergence sont d'ordre $\varepsilon^2$. (2) s'\'ecrit donc ${\rm div} \, [\xi_m \, U^2] = 0 (\varepsilon^2)$ ou encore
$$
\xi = \nabla W \times \frac{\vec U}{U^2} + 0 (\varepsilon^2) \qquad (\xi \cdot U \ \hbox{fait nul}) \, . \leqno (3)
$$
Pour faire le pas suivant, on reporte l'expression (3) dans la composante m\'eridienne de (1), qu'on va pouvoir r\'esoudre aux deux premiers ordres (l'ordre z\'ero est l'analogue de l'ordre ``cylindrique'' usuel, l'ordre un est celui des couplages de modes et aussi l'ordre o\`u joue la pression).

\smallskip

Cette r\'esolution va donner $\xi$ aux deux premiers ordres et en m\^eme temps $\phi$ aux deux premiers ordres. Bien entendu, \`a l'ordre z\'ero, on peut choisir \`a volont\'e le nombre d'ondes $m$ du mode qu'on calcule.

\smallskip

On peut par exemple \'eliminer $\phi$ en \'ecrivant :
$$
U \cdot {\rm rot} \, G = 0 (\varepsilon^2) \qquad \hbox{avec (3)}. \leqno (4)
$$
En n\'egligeant des quantit\'es d'ordre $\varepsilon^2$ on trouve :
$$
- U \cdot {\rm rot} \, G = {\rm div} \left[ U^2 \, \vec\nabla_m \left( \frac{B \nabla W}{U^2} \right)\right] + \nabla W \cdot {\rm rot} \, \vec J_u + \nabla p \cdot {\rm rot} \, [ \vec U (B \nabla)^{-1} (\nabla W \cdot {\rm rot})] \, .
\leqno (5)
$$

Il est aussi n\'ecessaire d'\'evaluer les quantit\'es qui interviennent dans les conditions de continuit\'e. Ce sont~:
\begin{eqnarray}
&&\xi \cdot \nabla F = \frac{B_m \nabla W}{U^2} \nonumber \\
&&\frac{B_m \cdot G}{U^2} = \frac{B_m \nabla \phi}{U^2} = -\nabla F \cdot \nabla \left( \frac{B \nabla W}{U^2} \right) + \frac{B_m \nabla W}{U^2} \, \frac{J \cdot U}{U^2} \, . \nonumber
\end{eqnarray}
La composante de $\frac{J \cdot U}{U^2}$ qui est \'egale \`a $2h\ell \, U^2$ ne joue pas \`a l'ordre principal dans l'\'equation d'Euler de $W$ parce que c'est une constante mais elle joue dans l'expression de $\phi$.

\smallskip

A l'ordre principal, apr\`es multiplication par $\frac{iq}{kf^2}$, on obtient :
$$
\frac{i}{kf} \, \frac{d\phi}{d\chi} = \left( \frac{i}{kq} \, \frac{d}{d\chi} - 1 \right) \rho \, \frac{dW}{d\rho} + 2 \left( \frac{h\ell}{k} - \frac{1}{qk} \right) i \, \frac{dW}{d\chi} \, .
$$
Pour l'harmonique $m$, en introduisant $\xi \cdot \nabla \mu = \frac{im}{\mu} \, W$
$$
A = \frac{i\phi}{\xi \cdot \nabla \mu} = kf \, \frac{\mu}{m} \left\{ \left( \frac{1}{kq} - \frac{1}{m} \right) \, \frac{\mu}{\xi \cdot \nabla \mu} \, \frac{d}{d\mu} \, (\xi \cdot \nabla \mu) + \frac{2h\ell}{k} - \frac{1}{kq} - \frac{1}{m} \right\} \, .
$$
Une fois r\'esolue l'\'equation de $W$, on revient \`a l'\'equation azimutale en y introduisant la valeur de $\phi$ qui est maintenant comme aux deux premiers ordres. On doit donc satisfaire l'\'equation de $U \cdot G$ avec un second membre, en tenant compte cette fois de tous les termes d'ordre $\varepsilon^2$, avec des valeurs approch\'ees de $\xi$, c'est-\`a-dire exprim\'ees en $W$. Autrement dit la correction porte seulement sur ${\rm div} (\xi_m \, U^2)$.

\smallskip

Dans le cas du kink interne, il est n\'ecessaire d'aller jusqu'\`a ce stade (2 ordres m\'eridiens, 3 ordres azimutaux) pour construire la fonction-test. Pour obtenir la vraie perturbation singuli\`ere, il faut m\^eme aller au 3$^{\rm e}$ ordre m\'eridien (selon le m\^eme principe it\'eratif). La formule (3) repr\'esente n\'ecessairement toutes les solutions aux deux premiers ordres. (C'est par exemple une cons\'equence \'evidente de $UDU=0$). En appliquant la formule $B_m \cdot D \cdot U = 0$ on obtient (pour la solution particuli\`ere)
$$
\xi \cdot B_m = \frac{B_m \nabla (\xi_u)}{U^2} + \frac{2h\ell}{k} \, U^2 \, i \, \xi \cdot \nabla F \, .
$$
Au deux premiers ordres on peut remplacer $i \, \xi \, \nabla F$ par $\frac{1}{U^2} \, B_m \, \nabla W$.

\bigskip

\begin{center}
\bf Annexe D -- Sens de rotation des diff\'erentes h\'elices
\end{center}

\medskip

\noindent 1) Pour d\'efinir une h\'elice on choisit un axe (orient\'e) $0z$, un rayon $r$ et une h\'elicit\'e (avec un signe). [Sens direct ou r\'etrograde ? (d\'efinir).] Sur un petit cercle de rayon $r$ et d'axe $0z$, on fixe une orientation (droite) pour l'angle ($\varphi$), de telle fa\c con que le tri\`edre $er$, $e\varphi$, $ez$ soit droit. On choisit un point arbitraire de l'h\'elice. L'h\'elicit\'e fixe le rapport constant de l'accroissement de $\varphi$ \`a celui de $z$ le long de l'h\'elice. Si elle est positive, l'h\'elice est droite, si elle est n\'egative, l'h\'elice est gauche.

\medskip

\noindent 2) L'h\'elice $(H)$ (l'axe magn\'etique) a pour h\'elicit\'e $\frac{h}{\ell}$. [Toute h\'elice de l'\'equilibre.] Puisque on prend $h$ toujours positif, $(H)$ est droite pour $\ell > 0$, gauche pour $\ell < 0$. Si $r,\varphi ,z$ est droit, $r,u,V$ est droit (le jacobien valant un). Sur $(H)$ $V = -\ell \varphi + hz$ est fix\'e.

\medskip

\noindent 3) Le syst\`eme intrins\`eque $\mu , u , \chi$ est droit ($\mu \, d \mu \, du \, d\chi = \frac{U}{V} \, dr \, du \, dV$). Pour les lignes magn\'etiques, le rayon est $\mu$, l'axe est $u$, l'angle est $\chi$. Il faut donc consid\'erer pour rep\'erer les h\'elices, le syst\`eme $\mu , \chi ,u$ qui est gauche. On a $u-q\chi =$ cte sur une ligne. Donc une ligne magn\'etique \`a $q$ positif est gauche. [La p\'eriode d'une ligne magn\'etique est $\Delta u = 2\pi q = \frac{2\pi}{k}$ longueur $2\pi \left( \frac{1}{Uk} \right)$ rapport d'aspect $k\rho$, $\frac{h\ell \rho}{\varepsilon}$.]

\medskip

\noindent 4) Une perturbation r\'esonante doit avoir l'h\'elicit\'e d'une ligne magn\'etique, donc \^etre gauche si $q$ est positif. Cela se traduit par le fait que $k$ a le m\^eme signe que $q$. Son facteur de p\'eriodicit\'e est d'ailleurs $e^{i (ku - q\chi)}$.

\medskip

\noindent 5) En conclusion, si $\frac{h\ell}{k}$ ou $h\ell q$ est positif, la perturbation a une h\'elicit\'e de signe oppos\'e \`a celle de l'axe ; si $\frac{h\ell}{k}$ est n\'egatif, les h\'elicit\'es sont de m\^eme signe.

\medskip

\noindent 6) De m\^eme pour les courants principaux $\vec U \, [2h\ell \, f \, U^2]$ et $\vec U \, V^2 \, LF$. Le second est \'egal \`a $- \, \vec U \, f \left[ \frac{1}{p} \, \frac{d}{d\rho} \, \frac{\rho^2}{q} \right] V^2$. Il est de sens oppos\'e au premier si $h \ell q > 0$.

\end{document}